\documentclass[a4paper]{jpconf}
\usepackage{graphicx}
\usepackage{lineno}
\usepackage{cite}

\def\pt{\mbox{$p_{\rm T}$ }}   
\def\pbpb {Pb--Pb }

\def\jpsi {\mbox{J/$\psi$ }}

\newcommand{\snn}  {\ensuremath{\sqrt{s_{\rm NN}}}}  

\begin{document}
\title{Charmonium production in Pb-Pb collisions at $\sqrt{s_{\rm NN}} =$ 2.76 and 5.02 TeV with ALICE}

\author{Biswarup Paul on behalf of the ALICE Collaboration}

\address{INFN Torino, Via Pietro Giuria 1, I-10125 Torino, Italy}

\ead{biswarup.paul@cern.ch}
%\linenumbers
\begin{abstract}

The production of charmonium states, J/$\psi$ and $\psi$(2S), in heavy-ion collisions, is an important probe to investigate the formation of a plasma of quarks and gluons (QGP). In a hot and deconfined medium, quarkonium production is, indeed, expected to be significantly modified, with respect to the pp yields, due to a balance of color screening and charm-quark recombination mechanisms. The ALICE Collaboration at the LHC has measured charmonium production in Pb-Pb collisions at $\sqrt{s}_{\rm NN} =$ 2.76 and 5.02 TeV. The nuclear modification factor of inclusive J/$\psi$, evaluated at forward ($-4<\eta<-2.5$) rapidity, is measured as a function of the centrality of the collision and of the J/$\psi$ kinematic variables as transverse momentum and rapidity. In this article, we report on the new J/$\psi$ results, obtained at forward rapidity, at $\sqrt{s}_{\rm NN} = 5.02$ TeV. These new results are compared to the J/$\psi$ nuclear modification factor obtained at $\sqrt{s}_{\rm NN} =$ 2.76 TeV and to the available theoretical predictions.

\end{abstract}

\section{Introduction}

The suppression of quarkonia (bound states of a heavy quark and its anti-quark) in ultra relativistic heavy-ion collisions is one of the most prominent probes used to investigate and quantify the properties of the QGP. The in-medium dissociation probability of the different quarkonium states could provide an estimate of the temperature of the fireball since the dissociation is expected to take place when the medium reaches or exceeds the critical temperature of the phase transition ($T_{\rm c}$), depending on the binding energy of the quarkonium state. For charmonium ({$\rm c \bar c$}) states, the J/$\psi$ is likely to survive significantly above $T_{\rm c}$ (1.5 -- 2 $T_{\rm c}$) whereas $\chi_{c}$ and $\psi$(2S) melt near $T_{\rm c}$ (1.1 -- 1.2 $T_{\rm c}$)~\cite{hsatz,hsatz2}. At LHC energies, due to the large increase of the c$\rm \bar c$ production cross-section with the collision energy, there is a possibility of J/$\psi$ production via recombination of deconfined c and $\rm \bar c$. Thus, the observation of J/$\psi$ production in nucleus-nucleus collisions via recombination also constitutes an evidence of QGP formation. The two effects act in opposite directions and the comparison of the J/$\psi$ production at the different energies can provide insights in the evolution of the relative contribution of the two processes.

%\section{ALICE detector and data samples}
The ALICE experiment has studied J/$\psi$ inclusive production in \pbpb\ collisions at $\sqrt{s}_{\rm NN} = \mbox{2.76 TeV}$ and 5.02 TeV through its dimuon decay channel. Muons are identified and tracked in the Muon Spectrometer, which covers the pseudorapidity range $-4<\eta<-2.5$~\cite{alice}. The pixel layers of the Inner Tracking System allow the vertex determination, while forward VZERO scintillators are used for triggering purposes. The VZERO is also used to determine the centrality of the collisions. In order to improve the purity of the muon tracks the following selection criteria were applied: (1) both muon tracks matched with trigger tracks above 1 GeV/$\it{c}$ \pt\ threshold, (2) both muon tracks in the pseudo-rapidity range $-$4 $<$ $\eta$ $<$ $-$2.5, (3) transverse radius coordinate of the tracks at the end of the hadron absorber (the longitudinal position of the absorber from the interation point (IP) is $-$5.0 $<$ $z$ $<$ $-$0.9 m) in the range 17.6 $<$ $R_{\rm{abs}}$ $<$ 89.5 cm and (4) dimuon rapidity in the range  2.5 $<$ $y$ $<$ 4. These cuts eliminate mainly tracks hitting the edges of the spectrometer’s acceptance or crossing the thicker part of the beam shield. The data samples used in this analysis correspond to an integrated luminosity $L_{\rm int}^{\rm {Pb-Pb}}\approx 225 \mu{\rm b}^{-1}$ for \pbpb\ at $\sqrt{s}_{\rm NN} =$ 5.02 TeV and $L_{\rm int}^{\rm {pp}}\approx 106\, {\rm nb}^{-1}$ for pp collisions at $\sqrt{s} =$ 5.02 TeV. The analysis techniques are described in detail in~\cite{pbpb5TeV}. 

\section{Results}

The in-medium modification of the J/$\psi$ production is quantified through the nuclear modification factor ($R_{\rm AA}$), defined as the ratio of the J/$\psi$ yield measured in Pb--Pb collisions and the expected yield obtained scaling the pp J/$\psi$ production cross section by the number of binary nucleon-nucleon collisions.
The $R_{\rm AA}$ of the inclusive J/$\psi$ in \pbpb collisions at $\sqrt{s_{\rm NN}}=\mbox{5.02 TeV}$, integrated over the centrality, rapidity and transverse momentum ($p_{\rm T}$) ranges (0--90\%, $2.5< y <4$ and $p_{\rm T}<12$ GeV/$c$) is $R_{\rm AA}=0.65\pm0.01 ({\rm stat}) \pm0.05 ({\rm syst})$~\cite{pbpb5TeV}. This shows that also at $\sqrt{s_{\rm NN}}=5.02$ TeV there is a suppression of the J/$\psi$ in Pb--Pb collisions with respect to pp collisions at the same energy.

\begin{figure}[ht]
\begin{center}
\includegraphics[width=7.5cm,height=5.5cm,angle=0]{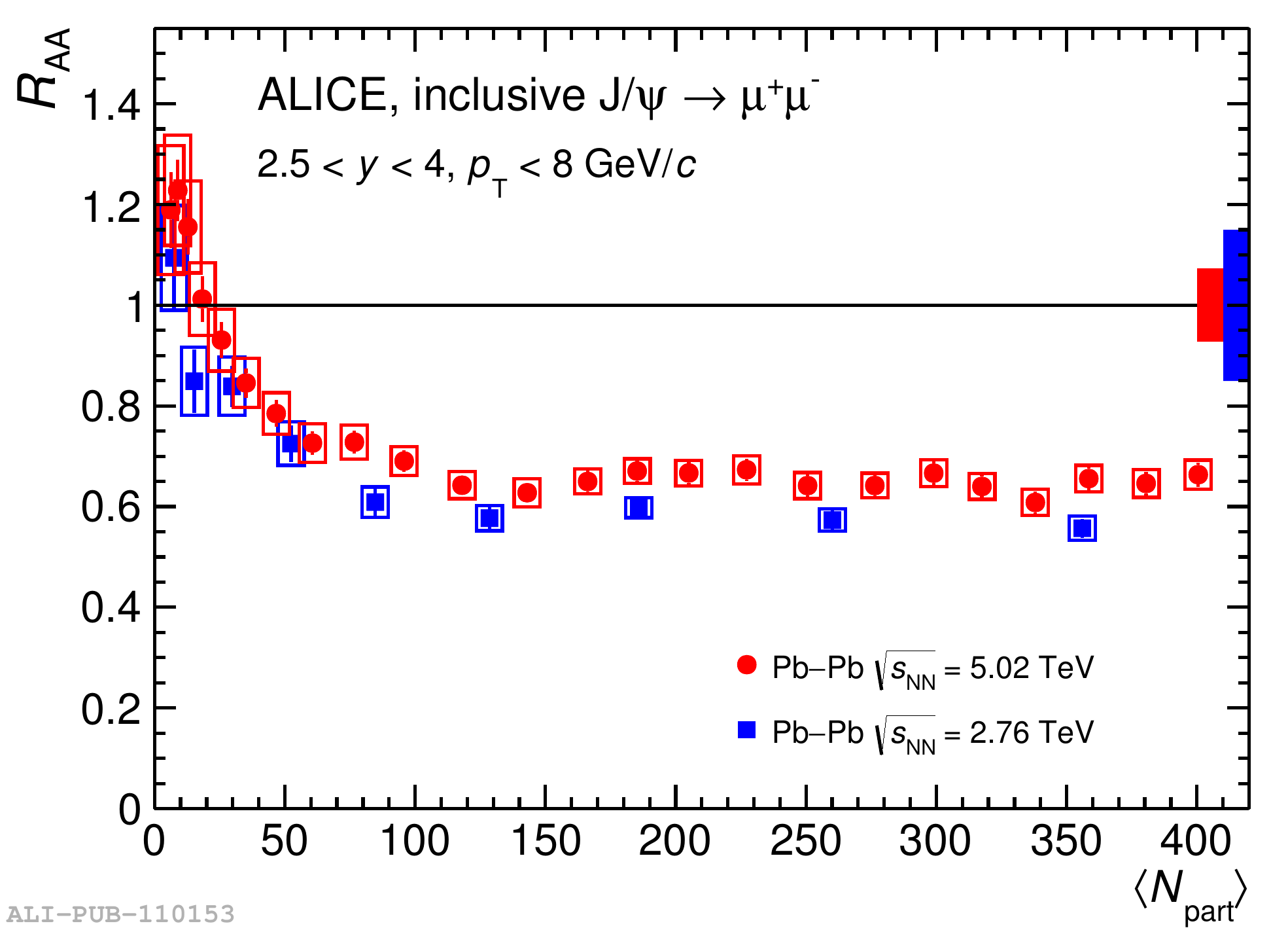}~~~~ \includegraphics[width=7.5cm,height=5.5cm,angle=0]{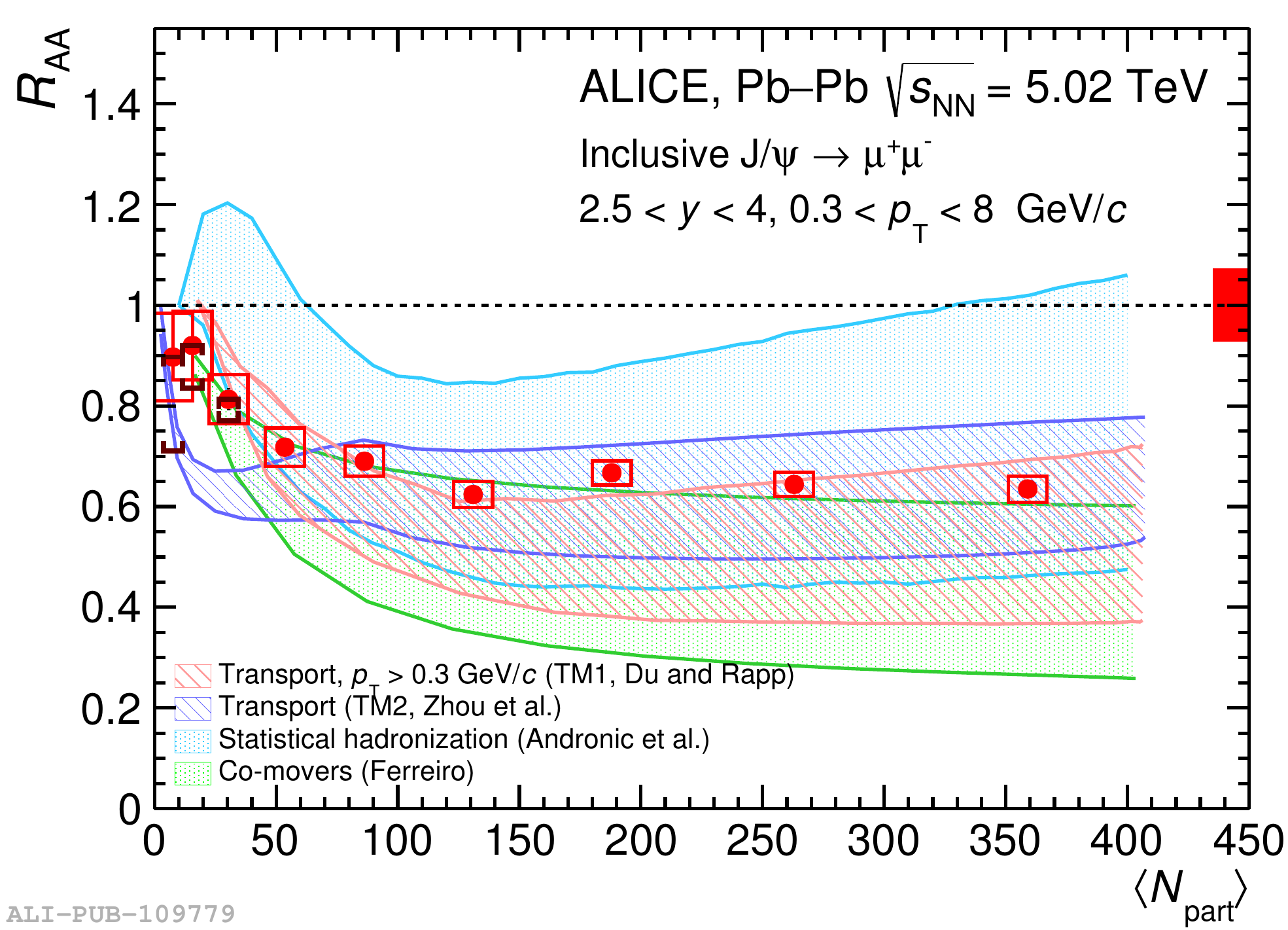}
\end{center}
\vspace{-3.500mm}
\caption{\small \small (Left) The inclusive J/$\psi$ $R_{\rm AA}$ as a function of centrality at $\sqrt{s_{\rm NN}}=5.02$ TeV compared to published results at $\sqrt{s_{\rm NN}}=2.76$ TeV~\cite{Abelev:2013ila}. The statistical uncertainties are represented by the error bars, the boxes around the points represent the uncorrelated systematic uncertainties, while correlated global uncertainties are shown as a filled box around $R_{\rm AA}=1$. (Right) The inclusive J/$\psi$ $R_{\rm AA}$ for $0.3<p_{\rm T}<\mbox{8 GeV/$c$}$ compared to the theoretical models~\cite{Andronic:2012dm,Ferreiro:2012rq,Ferreiro:2014bia,Zhao:2011cv,Du:2015wha,Zhou:2014kka}. }
\label{fig1}
\end{figure}

The centrality dependence of the inclusive J/$\psi$ $R_{\rm AA}$ at $\sqrt{s_{\rm NN}}=5.02$ TeV and the comparison with the corresponding values obtained at $\sqrt{s_{\rm NN}}=2.76$ TeV~\cite{Abelev:2013ila} is shown in Fig.~\ref{fig1}. The large statistics collected at $\sqrt{s_{\rm NN}}=5.02$ TeV allows to perform much narrower centrality bins w.r.t. to 2.76 TeV. The figure shows that there is an increasing suppression with centrality up to $N_{\rm part}\sim 100$ followed by a roughly constant $R_{\rm AA}$ value in both energies. A systematic increase by about 15\% is visible when comparing the two sets of results, although such an increase is within the total uncertainty of the measurements. The $R_{\rm AA}$ of prompt J/$\psi$ would be about 10\% higher if $R_{\rm AA}^{\rm non-prompt}=0$ (J/$\psi$'s produced from b-hadron decays are called non-prompt J/$\psi$) and about 5\% (1\%) smaller if $R_{\rm AA}^{\rm non-prompt}=1$ for central (peripheral) collisions. 

An excess of J/$\psi$ production at very-low $p_{\rm T}$, which might originate from photo-production, was observed in peripheral Pb--Pb collisions at $\sqrt{s_{\rm NN}}=2.76$ TeV~\cite{Adam:2015gba}. A $p_{\rm T}$ cut at 0.3 GeV/$c$ is applied in this analysis. According to the prediction from STARLIGHT MC simulation~\cite{photo-produced} this cut removes $\sim$80\% of the photo-produced J/$\psi$. The hadronic J/$\psi$ $R_{\rm AA}$, for $0 < p_{\rm T} < 8$ GeV/$c$, is estimated to be about 34\%, 17\% and 9\% smaller than the measured values in the 80--90\%, 70--80\% and 60--70\% centrality classes, respectively~\cite{Adam:2015isa}. The variation decreases to about 9\%, 4\% and 2\%, respectively, when considering the $R_{\rm AA}$ for J/$\psi$ with $0.3 < p_{\rm T} < 8$ GeV/$c$, due to the remaining small contribution of photo-produced J/$\psi$. Due to such selection a less pronounced increase of $R_{\rm AA}$ for peripheral events can indeed be observed as can be seen by comparing the left and right plots in Fig.~\ref{fig1}. The brackets shown in the right figure for the three most peripheral centrality intervals represent the range of variation of the hadronic J/$\psi$ $R_{\rm AA}$ under extreme hypothesis on the photo-production contamination on the inclusive $R_{\rm AA}$.

The results have been compared to the theoretical calculations based on a statistical model approach~\cite{Andronic:2012dm}, a transport model (TM1)~\cite{Zhao:2011cv,Du:2015wha}, which is based on a thermal rate equation, a second transport model (TM2)~\cite{Zhou:2014kka}, which implements a hydrodynamic description of the medium evolution and finally, or a `co-mover' model~\cite{Ferreiro:2012rq,Ferreiro:2014bia}, where the J/$\psi$ are dissociated via interactions with the partons/hadrons produced in the same rapidity range. The large uncertainties in the models are due to the choice of the corresponding input parameters and in particular of ${\rm d}\sigma_{\rm c\overline c}/{\rm d}y$. For most theoretical models a better description is found when considering their upper limit. But for transport models this corresponds to the absence of nuclear shadowing, which can be clearly considered as an extreme assumption for J/$\psi$~\cite{Adam:2015iga,Adam:2015jsa}.

\begin{figure}[ht]
\begin{center}
\includegraphics[width=7.5cm,height=5.5cm,angle=0]{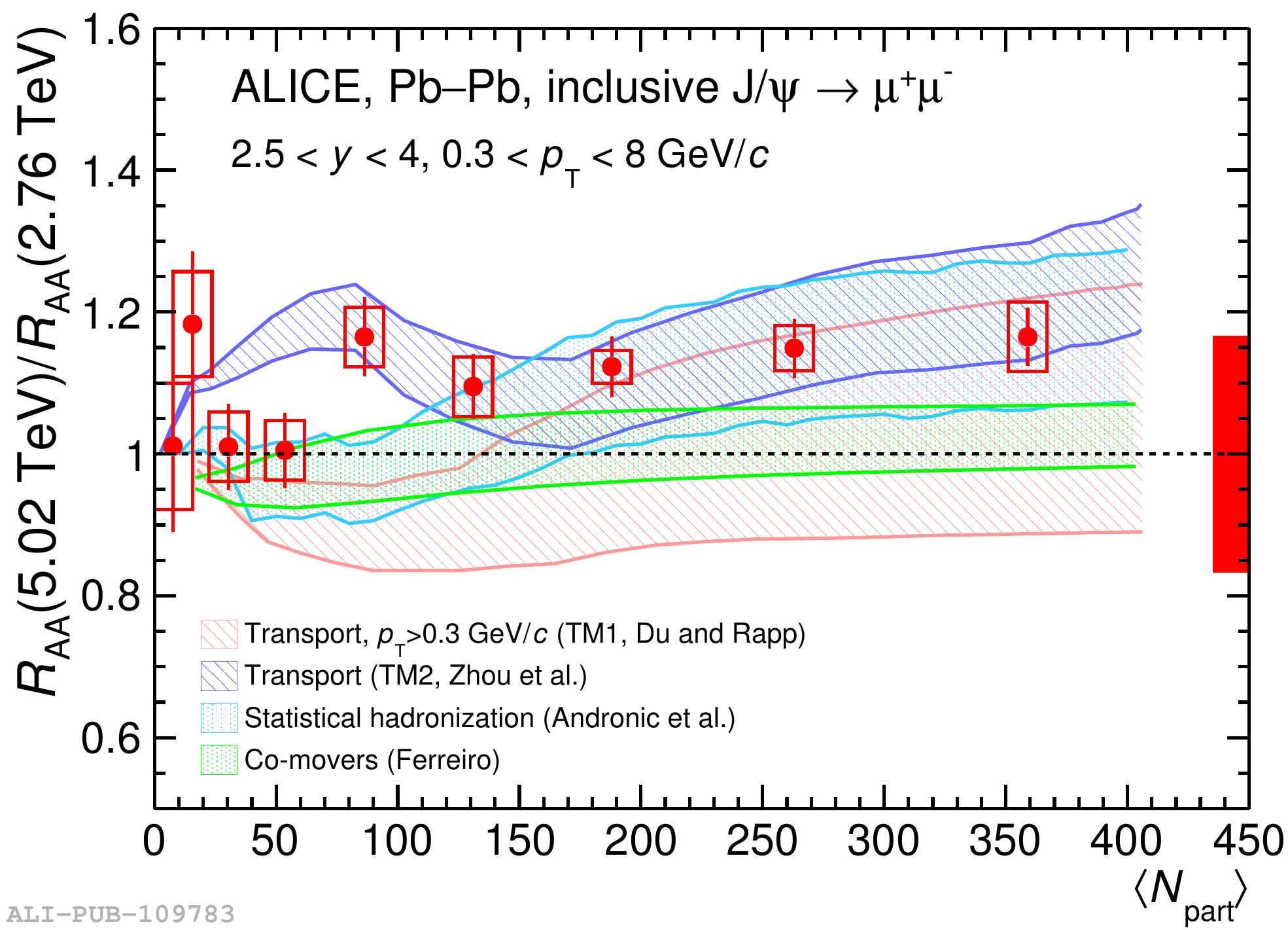}~~~~ \includegraphics[width=7.5cm,height=5.5cm,angle=0]{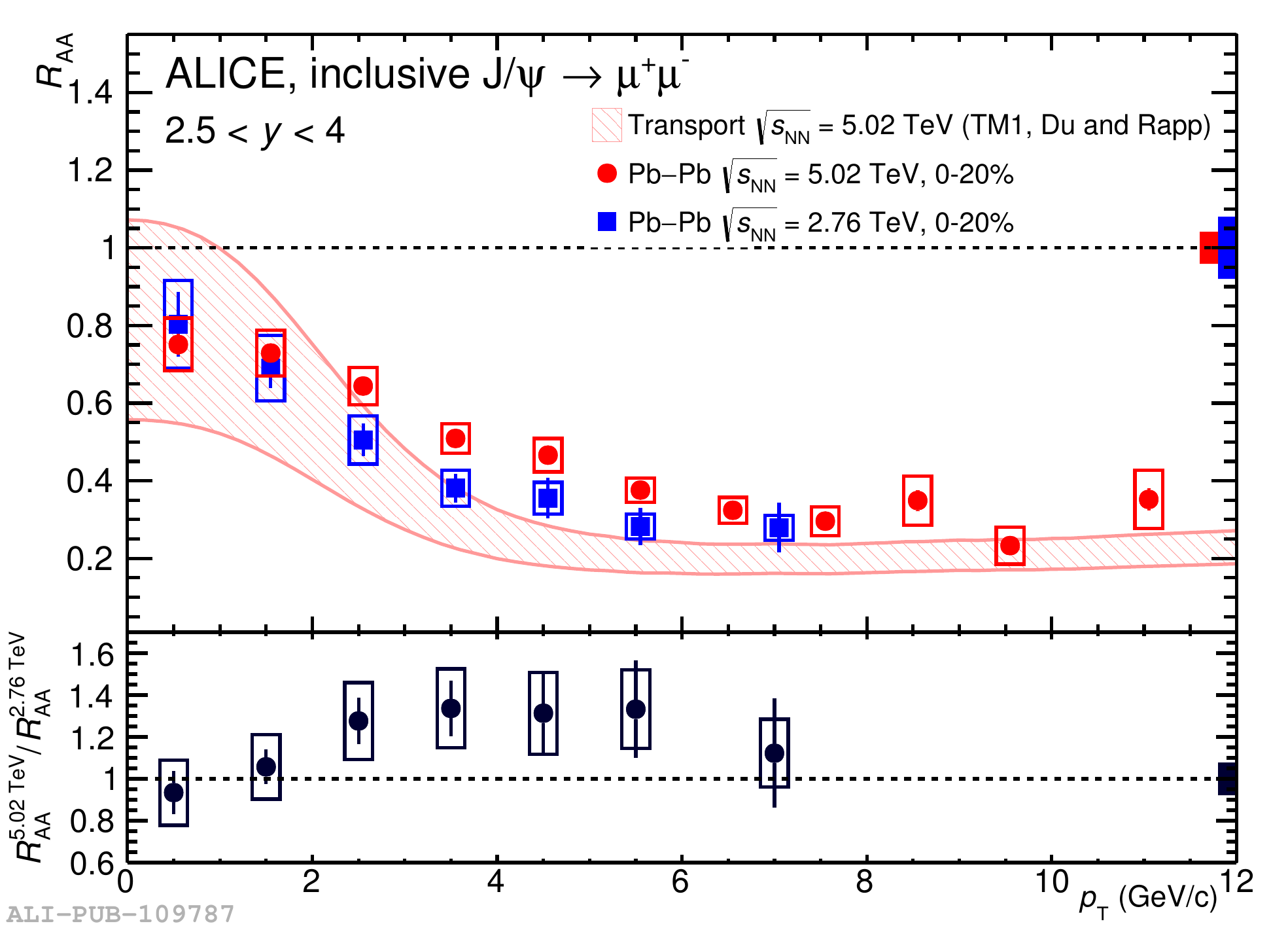}
\end{center}
\vspace{-3.500mm}
\caption{\small \small (Left) The ratio of the inclusive J/$\psi$ $R_{\rm AA}$ for $0.3<p_{\rm T}<8$ GeV/$c$ between $\sqrt{s_{\rm NN}}=5.02$ and 2.76 TeV vs centrality compared to theoretical models ~\cite{Andronic:2012dm,Ferreiro:2012rq,Ferreiro:2014bia,Zhao:2011cv,Du:2015wha,Zhou:2014kka}. (Right) The inclusive J/$\psi$ $R_{\rm AA}$ as a function of $p_{\rm T}$ at $\sqrt{s_{\rm NN}}=5.02$ TeV in the centrality interval 0--20\% compared to the corresponding result at $\sqrt{s_{\rm NN}}=2.76$ TeV~\cite{Abelev:2013ila} and to the prediction of a transport model~\cite{Zhao:2011cv,Du:2015wha}(TM1). The $p_{\rm T}$ dependence of the ratio of $R_{\rm AA}$ between the two energies is also shown.}
\label{fig2}
\end{figure}

The ratio of the inclusive J/$\psi$ $R_{\rm AA}$ between the two energies as a function of centrality and comparison with the theoretical models is shown in the left panel of Fig.~\ref{fig2}. Many theoretical uncertainties cancel out due to the correlation between the parameters at the two energies. The uncertainties on the average of the nuclear overlap function ($\langle T_{\rm AA}\rangle$) cancel out in data. This ratio for the prompt J/$\psi$ would be about 2\% (1--2\%) higher if beauty hadrons were fully (not) suppressed by the medium. The prediction from the transport model of Ref.~\cite{Zhao:2011cv,Du:2015wha} (TM1) shows a decrease of this ratio with the increase of centrality due to the larger suppression effects at high energy followed by an increase which is due to the regeneration that acts in the opposite direction and becomes dominant for central collisions. The other transport model (TM2)~\cite{Zhou:2014kka} also exhibits an increase for central collisions while for peripheral collisions the behaviour is different. No visible structure as a function of centrality is observed in the co-mover model~\cite{Ferreiro:2012rq,Ferreiro:2014bia} and the calculation favours this ratio to be slightly below unity which implies that in this model the increase of the suppression effects with energy may be dominant over the regeneration effects for all centralities. Finally, the statistical model~\cite{Andronic:2012dm} predicts a continuous increase of this ratio with centrality dominated by the increase in the ${\rm c}\overline{\rm c}$ cross section with energy. The uncertainty bands in the theoretical models shown in left panel of Fig.~\ref{fig2} correspond to $\sim$5\% variations of the $\rm c\overline c$ cross section at $\sqrt{s_{\rm NN}}=5.02$ TeV. Within the uncertainties the data show no clear centrality dependence and are compatible with the theoretical models. This ratio for central collisions and for $0.3<p_{\rm T}<8$ GeV/$c$ is $1.17 \pm 0.04 {\rm{(stat)}}\pm 0.20 {\rm{(syst)}}$. 

The $R_{\rm AA}$ as a function of $p_{\rm T}$ for the centrality interval 0--20\% compared to the corresponding results obtained at $\snn = 2.76$ TeV and to theoretical models is shown in the right panel of Fig.~\ref{fig2}. The $p_{\rm T}$ dependence of $R_{\rm AA}$ has an increasing trend at low $p_{\rm T}$, which is a sensitive test of the presence of a regeneration component in theoretical calculations. Since the contribution of J/$\psi$ photo-production is negligible with respect to the hadronic one for central events~\cite{Adam:2015gba}, the region $p_{\rm T}<0.3$ GeV/$c$ was not excluded. The transport model of Ref.~\cite{Zhao:2011cv,Du:2015wha} (TM1) describes the data at low $p_{\rm T}$ but underestimate the data at intermediate $p_{\rm T}$ ($3 < p_{\rm T} < 7$ GeV/$c$). The ratio of $R_{\rm AA}$ vs $p_{\rm T}$ between the two energies is also shown in the same figure. In the region $2<p_{\rm T}<6$ GeV/$c$, a hint of increase of $R_{\rm AA}$ with $\sqrt{s_{\rm NN}}$ is visible. The $R_{\rm AA}$ of prompt J/$\psi$ is expected to be 7\% larger (2\% smaller) for $p_{\rm T} < 1$ GeV/$c$ and 30\% larger (55\% smaller) for $10 < p_{\rm T} < 12$ GeV/$c$ when the beauty contribution is fully (not) suppressed. The ratio of $R_{\rm AA}$ appears to be less sensitive to the non-prompt J/$\psi$ contribution if we assume that $R_{\rm AA}^{\rm non-prompt}$ does not vary significantly between the two collision energies. 

\section{Conclusion}

ALICE has measured the inclusive \jpsi production in pp and \pbpb collisions at $\snn = 5.02$ TeV. Comparison with the previous measurement at $\snn=2.76$ TeV shows a systematic difference by about 15\%, although such an effect is within the total uncertainty of the measurements. A less pronounced increase of $R_{\rm AA}$ for peripheral events is observed upon removing the very-low $p_{\rm T}$ J/$\psi$ ($p_{\rm T}<0.3$ GeV/c). The rejection of photo-produced J/$\psi$~\cite{Adam:2015gba} is a possible explanation for this decrease. The theoretical calculations describe $R_{\rm AA}$ and the ratio of $R_{\rm AA}$ between $\sqrt{s_{\rm NN}}=5.02$ and 2.76 TeV closer to their upper limits. The $p_{\rm T}$ dependence of $R_{\rm AA}$ exhibits an increase at low $p_{\rm T}$ which is ascribed by the model to the contribution of regenerated J/$\psi$. In the transverse momentum region $2<p_{\rm T}<6$ GeV/$c$, a hint of an increase of $R_{\rm AA}$ between $\sqrt{s_{\rm NN}}=2.76$ and 5.02 TeV is observed. The results on J/$\psi$ production support a picture where a combination of suppression and regeneration takes place in the QGP, the two mechanisms being dominant at high and low $p_{\rm T}$, respectively.

\end{document}